\documentclass[prc,superscriptaddress,unsortedaddress,twocolumn,preprintnumbers]{revtex4}

\usepackage{amsmath, amssymb}

\def\fun#1#2{\lower3.6pt\vbox{\baselineskip0pt\lineskip.9pt
\def\dfrac#1#2{{\displaystyle\frac{#1}{#2}}}
\ialign{$\mathsurround=0pt#1\hfil##\hfil$\crcr#2\crcr\sim\crcr}}}
\newcommand{\beq}{\begin{equation}}
\newcommand{\eeq}{\end{equation}}
\newcommand{\bea}{\begin{eqnarray}}
\newcommand{\eea}{\end{eqnarray}}

\newcommand{\br}{\boldsymbol{r}}
\newcommand{\bsigma}{\boldsymbol{\sigma}}

\usepackage{graphicx}

\usepackage{times}
\usepackage{color}

\usepackage{ulem}
\usepackage{bm}

\renewcommand\sout{\bgroup\color{red} \ULdepth=-.5ex \ULset}

\begin{document}

\preprint{KUNS-2824 / NITEP 74}

\title{Proton induced deuteron knockout reaction as a probe of an isoscalar proton-neutron pair in nuclei}

\author{Yoshiki Chazono}
\email[]{chazono@rcnp.osaka-u.ac.jp}
\affiliation{Research Center for Nuclear Physics, Osaka University, 
Ibaraki 567-0047, Japan}

\author{Kenichi Yoshida}
\affiliation{Department of Physics, Kyoto University, 
Kyoto 606-8502, Japan}

\author{Kazuki Yoshida}
\affiliation{Advanced Science Research Center, Japan Atomic Energy Agency, 
Tokai, Ibaraki 319-1195, Japan}

\author{Kazuyuki Ogata}
\affiliation{Research Center for Nuclear Physics, Osaka University, 
Ibaraki 567-0047, Japan}
\affiliation{Department of Physics, Osaka City University, 
Osaka 558-8585, Japan}
\affiliation{Nambu Yoichiro Institute of Theoretical and Experimental Physics, 
Osaka City University, Osaka 558-8585, Japan}

\date{\today}

\begin{abstract}
\noindent
\textbf{Background:} The isoscalar $pn$ pair is expected to emerge in nuclei having the similar proton and neutron numbers but there is no clear experimental evidence for it. \\
\textbf{Purpose:} We aim to clarify the correspondence between the $pn$ pairing strength in many-body calculation and the triple differential cross section (TDX) of proton-induced deuteron knockout ($p,pd$) reaction on $^{16}$O. \\
\textbf{Methods:} The radial wave function of the isoscalar $pn$ pair with respect to the center of $^{16}$O is calculated with the energy density functional (EDF) approach and is implemented in the distorted wave impulse approximation (DWIA) framework. 
The $pn$ pairing strength $V_0$ in the EDF calculation is varied and the corresponding change in the TDX is investigated. \\
\textbf{Results:} A clear $V_0$ dependence of the TDX is found for the $^{16}$O($p,pd$)$^{14}$N($1_2^+$) at $101.3$~MeV. 
The nuclear distortion is found to make the $V_0$ dependence stronger. \\
\textbf{Conclusions:} Because of the clear $V_0$-TDX correspondence, the ($p,pd$) reaction will be a promising probe for the isoscalar $pn$ pair in nuclei. 
For quantitative discussion, further modification of the description of the reaction process will be necessary. 
\end{abstract}

\maketitle

\section{INTRODUCTION}
\label{sec1}
The nucleon-nucleon ($NN$) correlation is one of the most important properties to understand atomic nuclei. 
The pairing correlation of $pp$ and $nn$, for which the total isospin $T=1$ and total spin $S=0$, has extensively been studied for many years~\cite{Ber13,BB05}. 
Another type of $NN$ correlation is a spatially correlated two neutrons, i.e., dineutron, expected to emerge in a dilute system~\cite{Mig73}. 
After the invention of radioactive beams, properties of dineutron and how to probe it have been discussed theoretically and experimentally~\cite{Mat06,HS05,Nak06,Sim07,Kik16}. 
Development of the physics of unstable nuclei also provided a new opportunity to investigate $N \sim Z$ nuclei in medium- and heavy-mass regions; $N$ ($Z$) is the neutron (proton) number. 
In such nuclei, because the shell structure around the Fermi levels of $p$ and $n$ are similar to each other, the $pn$ correlation of either or both $T=0$ and $T=1$ types is expected to play an important role~\cite{FM14}. 
Recently, it was suggested with an energy density functional (EDF) approach that a $T=0$ $pn$ pairing vibrational mode possibly emerges in $N=Z$ nuclei~\cite{KenYoshi14,ELitivi18}. 
Among conceivable probes for the $T=0$ pairing inside $N \sim Z$ nuclei, we consider the deuteron knockout reaction for the transition. 

In this study, we discuss the proton-induced deuteron knockout reaction for $^{16}$O, $^{16}$O($p,pd$)$^{14}$N$^*$; $^{14}$N is in the $1^+_2$ state in the final channel. 
This reaction with $101.3$~MeV proton was carried out at Maryland~\cite{CSam82} and a triple differential cross section (TDX) of the same order of magnitude as that of $^{16}$O($p,2p$)$^{15}$N$_\textrm{g.s.}$ at $101.3$~MeV was obtained. 
This indicates that quite a large amount of $pn$ pair that is detected as deuteron may exist in $^{16}$O. 
In Ref.~\cite{CSam82}, a distorted wave impulse approximation (DWIA) calculation was performed with assuming a single-particle model for the bound deuteron and a deuteron spectroscopic factor was deduced. 
However, a more microscopic treatment of the $pn$ pair inside $^{16}$O will be important to clarify its correspondence to the TDX of the $^{16}$O($p,pd$)$^{14}$N$^*$ reaction. 

To achieve this, we adopt the EDF for describing the structure of $^{16}$O, i.e., the radial wave function of the $pn$ pair regarding the center of $^{16}$O. 
The DWIA calculation is then performed to evaluate the TDX. 
Our main purpose is to clarify how the TDX behaves when the $pn$ pairing strength $V_0$ is changed in the EDF calculation. 
The distortion effect on the TDX-$V_0$ correspondence is discussed as well as the spatial region of $^{16}$O that is relevant to the ($p,pd$) process. 

The construction of this paper is as follows. 
In Sec.~\ref{sec2} we present the formalism of DWIA and EDF for the calculation of the TDX of $^{16}$O($p,pd$)$^{14}$N$^*$. 
We show numerical results of the structural calculation and the TDX in Sec.~\ref{sec3}. 
A summary and perspective are given in Sec.~\ref{sec4}.

\section{FORMALISM}
\label{sec2}

\subsection{DWIA framework}
\label{subsec2A}
We consider the $^{16}$O($p,pd$)$^{14}$N$^*$ reaction in normal kinematics; in the final channel $^{14}$N is assumed to be in the second $1^+$ excited state. 
The incoming proton is labeled as particle 0, and the outgoing proton and deuteron are labeled as particles 1 and 2, respectively. 
We denote the target (residual) nucleus $^{16}$O ($^{14}$N$^*$) by A (B) and its mass number by $A$ ($B$). 
In what follows, $\hbar \bm{K}_i$, $E_i$, and $T_i$ represent the momentum, the total energy, and the kinetic energy of particle $i$ ($=0,1,2,\textrm{A},\textrm{or B}$), respectively. 
The solid angle of the outgoing particle $j$ ($=1\textrm{ or }2$) is denoted by $\Omega_j$. 
The quantities with and without the superscript L represent that we evaluate these in the laboratory (L) and $p$-A center-of-mass (c.m.) frames, respectively. 

\begin{figure}[htbp]
\begin{center}
\includegraphics[width=0.45\textwidth]{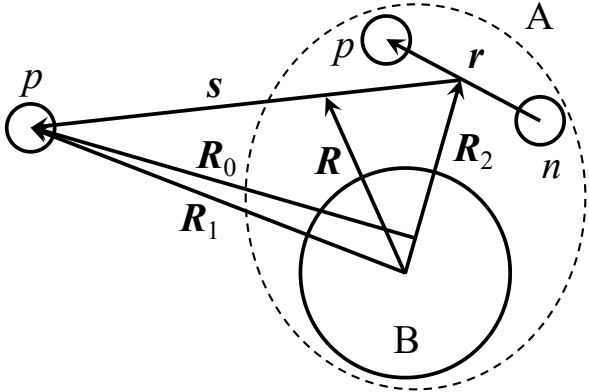}
\caption{Coordinates of the A($p,pd$)B reaction system.\label{fig1}}
\end{center}
\end{figure}

In the distorted wave impulse approximation (DWIA) framework, the transition amplitude of the A($p,pd$)B reaction is given by 
\begin{align}
T
&=
\left\langle \chi^{(-)}_{1, \bm{K}_1} (\bm{R}_1) 
\phi_d (\bm{r}) \chi^{(-)}_{2, \bm{K}_2} (\bm{R}_2) \right| \nonumber \\
&\qquad \times 
t_{pd} (\bm{s}) 
\left| \chi^{(+)}_{0, \bm{K}_0} (\bm{R}_0) 
\phi_d (\bm{r}) \varphi_{pn} (\bm{R}_2) \right\rangle, 
\label{eq1}
\end{align}
where $\chi_{i, \bm{K}_i}$ with $i=0,1,\textrm{ and }2$ are the distorted waves of the $p$-A, $p$-B, and $d$-B systems, respectively. 
The coordinate between the incoming (outgoing) proton and A (B) is denoted by $\bm{R}_0$ ($\bm{R}_1$) and that between the outgoing deuteron and B by $\bm{R}_2$. 
As seen from Fig.~\ref{fig1}, $\bm{R}_2$ also means the coordinate of the c.m.~of the isoscalar ($T=0$) spin-triplet ($S=1$) $pn$ pair relative to B inside A. 
The scattering waves with the superscripts $(+)$ and $(-)$ satisfy the outgoing and incoming boundary conditions, respectively. 
$\phi_d$ is the $pn$ relative wave function in the ground state of deuteron and 
$t_{pd}$ is the effective interaction between $p$ and $d$. 
The coordinates relevant to $\phi_d$ and $t_{pd}$ are denoted by $\bm{r}$ and $\bm{s}$, respectively. 
$\varphi_{pn}$ defined by 
\begin{align}
\varphi_{pn} (\bm{R}_2) = 
\left\langle \Psi_\textrm{B} \right| 
\left. \Psi_\textrm{A} \right\rangle_{\xi_\textrm{B}} 
\label{eq2}
\end{align}
is the wave function between the c.m.~of the $pn$ pair and B inside A; $\Psi_\textrm{C}$ ($\textrm{C}=\textrm{A or B}$) is the many-body wave function of C. 
In Eq.~\eqref{eq2}, it is understood that the integration is taken over all the intrinsic coordinates $\xi_\textrm{B}$ of B. 
A detailed description of $\varphi_{pn}$ is given in Sec.~\ref{subsec2B}. 

We apply the asymptotic momentum approximation~\cite{KazuYoshi16} to the distorted waves in Eq.~\eqref{eq1} and obtain 
\begin{align}
T \approx 
\tilde{t}_{pd} (\bm{\kappa}', \bm{\kappa}) 
\int d \bm{R}~ F (\bm{R}) \varphi_{pn} (\bm{R}). 
\label{eq3}
\end{align}
Here, $\bm{\kappa}$ ($\bm{\kappa}'$) indicates the relative momentum between $p$ and $d$ in the initial (final) state, and we define $\tilde{t}_{pd}$ and $F$ as follows: 
\begin{align}
\tilde{t}_{pd} (\bm{\kappa}', \bm{\kappa}) \equiv 
\left\langle \phi_d (\bm{r}) e^{i \bm{\kappa}' \cdot \bm{s}} \right| 
t_{pd} (\bm{s}) 
\left| \phi_d (\bm{r}) e^{i \bm{\kappa} \cdot \bm{s}} \right\rangle, 
\label{eq4}
\end{align}
\begin{align}
F (\bm{R}) \equiv 
\chi^{* (-)}_{1, \bm{K}_1} (\bm{R}) \chi^{* (-)}_{2, \bm{K}_2} (\bm{R}) 
\chi^{(+)}_{0, \bm{K}_0} (\bm{R}) e^{- 2 i \bm{K}_0 \cdot \bm{R} / A}. 
\label{eq5}
\end{align}

Using the final-state on-the-energy-shell prescription, i.e., 
\begin{align}
\bm{\kappa} \approx \kappa' \hat{\bm{\kappa}}, 
\label{eq6}
\end{align}
in the evaluation of $\tilde{t}_{pd}$, we find 
\begin{align}
\frac{\mu^2_{pd}}{(2 \pi \hbar^2)^2} \frac{1}{6} 
\left| \tilde{t}_{pd} (\bm{\kappa}', \bm{\kappa}) \right|^2 
\approx 
\frac{d \sigma_{pd}}{d \Omega_{pd}} (\theta_{pd}, E_{pd}), 
\label{eq7}
\end{align}
where $d \sigma_{pd} / d \Omega_{pd}$ is the $p$-$d$ elastic differential cross section in free space with $\theta_{pd}$ and $E_{pd}$ being the c.m.~scattering angle and the scattering energy, respectively. 
$\mu_{pd}$ is the reduced mass of the $p$-$d$ system. 

The triple differential cross section (TDX) for the A($p,pd$)B reaction is then given by 
\begin{align}
\frac{d^3 \sigma}{d E^\textrm{L}_1 d \Omega^\textrm{L}_1 d \Omega^\textrm{L}_2} = 
F_\textrm{kin} C_0 \frac{d \sigma_{pd}}{d \Omega_{pd}} (\theta_{pd}, E_{pd}) 
\left| \bar{T} \right|^2, 
\label{eq8}
\end{align}
where 
\begin{align}
F_{\rm kin} \equiv 
J_\textrm{L} \frac{K_1 K_2 E_1 E_2}{(\hbar c)^2} 
\left[ 1 + \frac{E_2}{E_\textrm{B}} + 
\frac{E_2}{E_\textrm{B}} \frac{\bm{K}_1 \cdot \bm{K}_2}{K_2} \right]^{-1}, 
\label{eq9}
\end{align}
\begin{align}
C_0 \equiv 
\frac{E_0}{(\hbar c)^2 K_0} 
\frac{\hbar^4}{(2 \pi)^3 \mu^2_{pd}}, 
\label{eq10}
\end{align}
and 
\begin{align}
\bar{T} \equiv 
\int d \bm{R}~ F (\bm{R}) \varphi_{pn} (\bm{R}). 
\label{eq11}
\end{align}
In Eq.~\eqref{eq9}, $J_\textrm{L}$ is the Jacobian from the $p$-A c.m.~frame to the L frame. 

Once all the distorting potentials are switched off, i.e., the plane wave impulse approximation (PWIA) is adopted, $\bar{T}$ turns out to be the Fourier transform of $\varphi_{pn} (\bm{R})$: 
\begin{align}
\bar{T} \approx 
\int d \bm{R}~ e^{- i \bm{K}_{pn} \cdot \bm{R}} \varphi_{pn} (\bm{R}), 
\label{eq12}
\end{align}
where $\bm{K}_{pn}$ is given by 
\begin{align}
\bm{K}_{pn} = 
\bm{K}_1 + \bm{K}_2 - \left( 1 - \frac{2}{A} \right) \bm{K}_0. 
\label{eq13}
\end{align}
By assuming the residual nucleus B is a spectator, one can interpret $\bm{K}_{pn}$ as the momentum of the c.m.~of the $pn$ pair. 

In the recoilless (RL) condition, which is characterized by $K_{pn}=0$, one finds 
\begin{align}
\bar{T} \approx 
\int d \bm{R}~\varphi_{pn} (\bm{R}) \equiv 
\mathcal{A}_{pn}. 
\label{eq14}
\end{align}
This clearly shows that the TDX in the RL condition reflects the total amplitude $\mathcal{A}_{pn}$ of the $pn$ pair.

\subsection{Microscopic calculation of the $pn$ pair wave function}
\label{subsec2B}
We apply the nuclear energy-density functional (EDF) method to describing microscopically the wave function of the $pn$ pair.
In a framework of the nuclear EDF, the $pn$-pair-removed excited states in $^{14}$N are described in the proton-neutron hole-hole Random-Phase Approximation (pn-hhRPA)~\cite{KenYoshi14} considering the ground-state of $^{16}$O as an RPA vacuum; 
$\displaystyle |\Psi_\textrm{B} \rangle = \Gamma^\dagger |\Psi_\textrm{A} \rangle$, 
where $\hat{\Gamma}^\dagger$ represents the RPA phonon operator; 
\begin{align}
\hat{\Gamma}^\dagger = 
\sum_{i i'} X_{i i'} \hat{b}^\dagger_{p,i} \hat{b}^\dagger_{n,i'} - 
\sum_{m m'} Y_{m m'} \hat{b}^\dagger_{n,m'} \hat{b}^\dagger_{p,m}. 
\label{eq15}
\end{align}
Here, $\hat{b}^\dagger_{p,i}$ ($\hat{b}^\dagger_{n,i'}$) create a proton (neutron) hole in the single-particle level $i$ ($i'$) below the Fermi level, and $\hat{b}^\dagger_{p,m}$ ($\hat{b}^\dagger_{n,m'}$) create a proton (neutron) hole above the Fermi level. 
Note that the backward-going amplitudes $Y$ vanish if the ground-state correlation in $^{16}$O is neglected. 
The single-particle basis is obtained as a self-consistent solution of the Skyrme-Hartree-Fock (SHF) equation. 

\begin{widetext}
The $S=1$ $pn$-pair-removal transition density that we need for the transition amplitude is given as 
\begin{align}
\delta \bar{\rho}_{\mu} (\br_n, \br_p) = 
\frac{1}{2} \sum_{\sigma \sigma'}(-2 \sigma') 
\left\langle \sigma' | \bsigma_\mu | \sigma \right\rangle 
\left\langle \Psi_\textrm{B} | 
\hat{\psi}_n (\br_n -\!\sigma') \hat{\psi}_p (\br_p \sigma) - 
\hat{\psi}_p (\br_p -\!\sigma') \hat{\psi}_n (\br_n \sigma) 
| \Psi_\textrm{A} \right\rangle, 
\label{eq16}
\end{align}
\end{widetext}
where $\bsigma = (\sigma_{-1}, \sigma_0, \sigma_{+1})$ denotes the spherical components of the Pauli spin matrices, and $\hat{\psi}_q (\br \sigma)$ the nucleon annihilation operator at the position $\br$ with the spin direction $\sigma = \pm 1/2$ expanded in the single-particle basis with $q=n$ or $p$. 
Since the transition density is spherical in spin space, we have only to consider one of the components for $\mu$. 
Here, we take the $\mu=0$ component of the wave function. 

From Eqs.~\eqref{eq1}, \eqref{eq2}, and \eqref{eq16}, we can regard the transition density $\delta \bar{\rho}_{0}$ as 
\begin{align}
\delta \bar{\rho}_{0} (\bar{\bm{R}}, \bm{r}) \approx 
\varphi_{pn} (\bm{R}) \phi_d (\bm{r}), 
\label{eq17}
\end{align}
where $\bar{\bm{R}} = (\br_n + \br_p)/2$ and $\br = \br_p - \br_n$ and 
\begin{align}
\varphi_{pn} (\bm{R}) \equiv 
\frac{\delta \bar{\rho}_{0} (\bm{R}, 0)}{\phi_d (0)} = 
\hat{\varphi}_{pn} (R) Y_{00} (\Omega_{\bm R}). 
\label{eq18}
\end{align}
Thus, in evaluating $\varphi_{pn}$, we consider the $pn$ pair is $S$-wave and a point particle, namely $\br=0$. 
The use of Eqs.~\eqref{eq17} and \eqref{eq18} means that the component of the $^{16}$O wave function that contains a deuteron is selected out. 
This treatment is consistent with the DWIA framework described in Sec.~\ref{subsec2A}. 

With this, the $pn$-pair-removal transition strength is given by 
\begin{align}
3 \left| 4 \pi \int^\infty_0 d R~ 
R^2 \hat{\varphi}_{pn} (R) \phi_d (0) \right|^2, 
\label{eq19}
\end{align}
where the factor three comes from the sum of $\mu=-1$, 0, and 1 components.

\section{RESULTS AND DISCUSSION}
\label{sec3}

\subsection{Numerical inputs}
\label{subsec3A}
To obtain the single-particle basis used in the pn-hhRPA calculation, the SHF equation is solved in cylindrical coordinates $\br = (r, z, \phi)$ with a mesh size of $\Delta r = \Delta z = 0.6$~fm and with a box boundary condition at $(r_\textrm{max}, z_\textrm{max}) = (14.7, 14.4)$~fm. 
The axial and reflection symmetries are assumed in the ground state, and the ground-state of $^{16}$O is calculated to be spherical. 
More details of the calculation scheme are given in Ref.~\cite{KenYoshi13}. 
The SGII interaction~\cite{NVGiai81} is used for the particle-hole (ph) channel, and the density-dependent contact interaction defined by 
\begin{align}
v_\textrm{pp} &(\br \sigma \tau, \br' \sigma' \tau') = \nonumber \\
&V_0 \frac{1 + P_\sigma}{2} \frac{1 - P_\tau}{2} 
\left[ 1 - \frac{\rho (\br)}{\rho_0} \right] \delta (\br - \br') 
\label{eq20}
\end{align}
is employed for the particle-particle (pp) channel. 
Here, $\rho_0$ is $0.16$~fm$^{-3}$ and $\rho (\br) = \rho_p (\br) + \rho_n (\br)$. 
We adopt three values $-100$, $-490$, and $-600$~MeV fm$^3$ for the pairing strength $V_0$. 

For the distorting potentials of proton, the EDAD1 parameter set of the Dirac phenomenology~\cite{SHama90,EDCoop93} is used, whereas we employ the global optical potential by An and Cai~\cite{HAn06} for deuteron. 
We construct the Coulomb potential in each distorting potential by assuming a uniformly charged sphere with the radii of $r_0 C^{1/3}~(C = A \textrm{ or } B)$ with $r_0$ being $1.41$~fm. 
Nonlocality corrections to the distorted waves of deuteron and proton are made by multiplying the wave functions by the Perey factor~\cite{Per63} with the 0.54~fm of the range of nonlocality and the Darwin factor~\cite{SHama90,Arn81}, respectively. 
For the $p$-$d$ elastic cross section in Eq.~\eqref{eq8}, 
we take the experimental data from Refs.~\cite{MDavi63, CCKim64, KKuro64, FHin68, SNBun68, TACahi71, NEBooth71, HShimi82, KSaga94, KSeki02, KHata02, KErm05} with the Lagrange interpolation with respect to the scattering angle and energy. 
The kinematics of all the particles are treated relativistically. 
The M\o ller factor~\cite{Mol45,Ker59} is taken into account to describe the transformation of the $p$-$d$ transition matrix from the $p$-$d$ c.m. frame to the $p$-A c.m. frame.

\subsection{Structure of the low-lying $1^+$ states in $^{14}$N}
\label{subsec3B}

\begin{figure}[htbp]
\begin{center}
\includegraphics[width=0.45\textwidth]{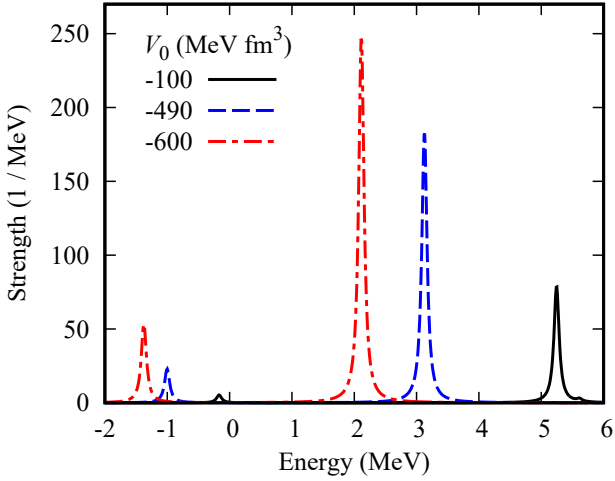}
\caption{$S=1$ $pn$-pair removal transition strengths as functions of the excitation energy, where the excitation energy for the unperturbed $(p_{1/2})^{-2}$ configuration is set as zero. 
The solid, dashed, and dot-dashed lines correspond to the cases of $V_0 = -100$, $-490$, and $-600$~MeV~fm$^3$, respectively.\label{fig2}}
\end{center}
\end{figure}

We briefly mention the structure of the calculated low-lying $1^+$ states in $^{14}$N in the present framework before discussing the TDX. 
Figure~\ref{fig2} shows the $S=1$ $pn$-pair-removal transition-strength distributions. 
The excitation energy is defined with respect to the excitation energy of the simplest configuration of $(p_{1/2})^{-2}$ coupled to $T=0, S=1$ in $^{16}$O. 
The lowest state and the second lowest state for each pairing strength correspond to the ground $1^+$ state and the $1^+_2$ state that we are interested in, respectively. 
They are constructed by mainly the $(\nu p_{1/2})^{-1} (\pi p_{1/2})^{-1}$ configuration, and the superposition of the $(\nu p_{1/2})^{-1} (\pi p_{3/2})^{-1}$ and $(\nu p_{3/2})^{-1} (\pi p_{1/2})^{-1}$ configurations, respectively. 
With an increase of the pairing strength, the energies become lower and the strengths get enhanced for both states. 
For the case of $V_0 = -600$~MeV~fm$^3$, the $(p_{3/2})^{-2}$ configuration is not negligible for enhancing the transition strength to the $1^+_2$ state. 
Therefore, the collectivity of the $1^+_2$ state becomes stronger with an increased pairing strength. 

The question arisen here is how much of the pairing strength we should employ. 
We are going to look at the energy difference of the $1^+$ states; $\Delta E = E_{1^+_2}-E_{1^+_1}$. 
For the case of $V_0=-100$, $-490$, and $-600$~MeV~fm$^3$, 
the calculated $\Delta E$ is $5.41$, $4.12$, and $3.48$~MeV, respectively, while $\Delta E = 3.95$~MeV experimentally. 
We can thus say that the pairing strengths $V_0 = -490$ and $-600$~MeV~fm$^3$ are a reasonable choice in the present study. 

\begin{figure}[htbp]
\begin{center}
\includegraphics[width=0.45\textwidth]{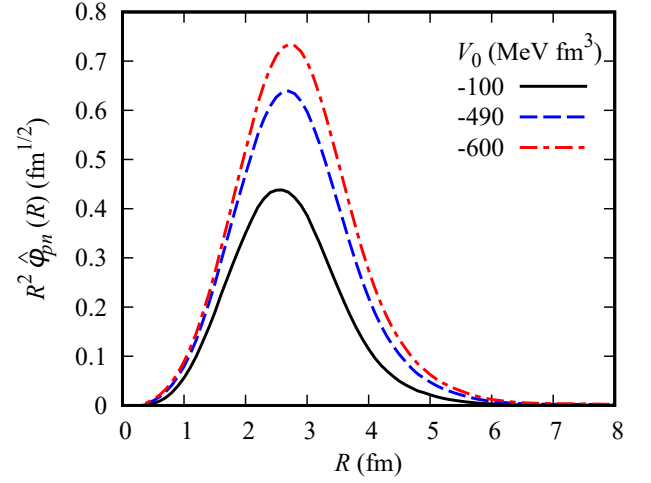}
\caption{Radial component of $\varphi_{pn} (\bm{R})$. 
The solid, dashed, and dot-dashed lines correspond to the cases of $V_0 = -100$, $-490$, and $-600$~MeV~fm$^3$, respectively. 
Note that each line is multiplied by $R^2$.\label{fig3}}
\end{center}
\end{figure}

Next, we check the behavior of $\varphi_{pn} (\bm{R})$. 
The radial components of $\varphi_{pn} (\bm{R})$ with $V_0 = -100$~MeV~fm$^3$ (solid line), $-490$~MeV~fm$^3$ (dashed line), and $-600$~MeV~fm$^3$ (dot-dashed line), respectively, are shown in Fig.~\ref{fig3}. 
It should be noted that each line is multiplied by $R^2$. 
One can clearly find that the stronger the pair interaction is, 
the larger the amplitude of $R^2 \hat{\varphi}_{pn} (R)$ is, i.e., 
the stronger collectivity the $pn$ pair has. 
Note that in the $V_0 \rightarrow 0$ limit, the independent-particle picture of $^{16}$O is realized. Then, the peak of the $ R^2 \hat{\varphi}_{pn} (R)$ will almost disappear.

\subsection{Triple differential cross section for $^{16}$O(\textit{p},\textit{pd})$^{14}$N$^*$ reaction at 101.3~MeV}
\label{subsec3C}

\begin{figure}[htbp]
\begin{center}
\includegraphics[width=0.45\textwidth]{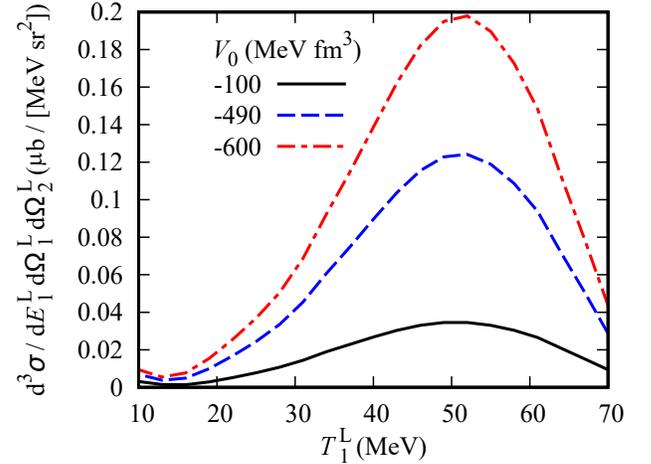}
\caption{Triple differential cross section (TDX) for the $^{16}$O($p,pd$)$^{14}$N$^*$ reaction at $101.3$~MeV. 
The solid, dashed, and dot-dashed lines corresponds the results with $\hat{\varphi}_{pn}$ of $V_0 = -100$, $-490$, and $-600$~MeV~fm$^3$, respectively.\label{fig4}}
\end{center}
\end{figure}

In Fig.~\ref{fig4} we show the TDX for the $^{16}$O($p,pd$)$^{14}$N$^*$ reaction at $101.3$~MeV as a function of $T^\textrm{L}_1$. 
The emission angle of particle 1 is fixed at $(\theta^\textrm{L}_1, \phi^\textrm{L}_1) = (40.1^\circ, 0^\circ)$ and that for particle 2 at $(\theta^\textrm{L}_2, \phi^\textrm{L}_2) = (40.0^\circ, 180^\circ)$; we follow the Madison convention. 
At $T^\textrm{L}_1 \sim 52$~MeV, the RL condition is almost satisfied. 
This kinematical condition corresponds to $E_{pd} \sim 56$~MeV and $\theta_{pd} \sim 68^\circ$ for the $p$-$d$ scattering. 
The results using $\hat{\varphi}_{pn}$ calculated with $V_0 = -100$, $-490$, and $-600$~MeV~fm$^3$ are shown by the solid, dashed, and dot-dashed lines, respectively. 
One sees a clear correspondence between $V_0$ and the TDX. 
In other words, the height of the TDX reflects the collectivity of the $pn$ pair that forms deuteron in $^{16}$O. 
Unfortunately, however, it is difficult to make a quantitative comparison of the current results with experimental data. 
This is mainly because of the approximate treatment of $\varphi_{pn}$ in Eq.~\eqref{eq18}; the TDX of knockout reactions is known to be quite sensitive to the radial distribution of the wave function of the particle to be knocked out, which may be affected by the approximation of Eq.~\eqref{eq18} in the present case. 
A sensitivity test of the TDX on $\varphi_{pn}$ is given in Appendix. 
Besides, there may exist other reaction mechanisms that are not considered in this study; we come to this point in Sec.~\ref{sec4}. 
Nevertheless, the $V_0$ dependence of the TDX can safely be investigated, which is our primary objective of this study. 

\begin{figure}[htbp]
\begin{center}
\includegraphics[width=0.45\textwidth]{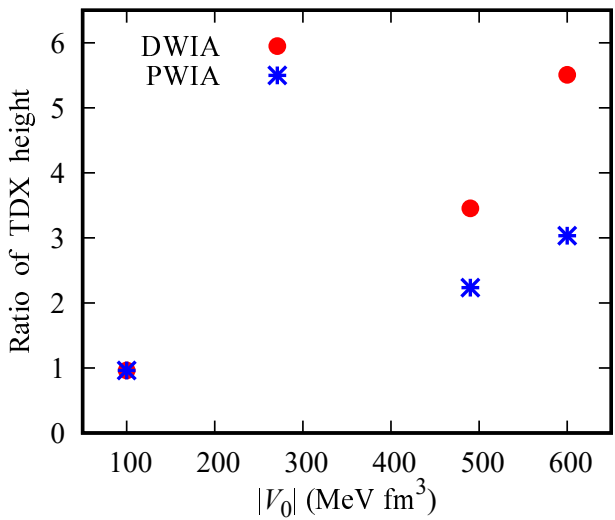}
\caption{Ratio of the TDX height to that calculated with $V_0 = -100$~MeV fm$^3$. 
The circles (asterisks) represent the results of the DWIA (PWIA) calculations.\label{fig5}}
\end{center}
\end{figure}

To see the $V_0$-TDX correspondence more clearly, in Fig.~\ref{fig5} we show the values of the TDX at $T^\textrm{L}_1 = 52$~MeV, the TDX height, in ratio to the value calculated with $V_0 = -100$~MeV~fm$^3$. 
The result of the DWIA (PWIA) is represented by the circles (asterisks). 
As mentioned above, $T^\textrm{L}_1 = 52$~MeV corresponds to the RL condition. 
In the PWIA limit, one expects from Eqs.~\eqref{eq8} and \eqref{eq14} a clear relation between the TDX height and $|\mathcal{A}_{pn}|^2$, as shown by the asterisks. 
When the distortion is included, the ratio is found to increase further. 
This indicates that the TDX height observed in the $^{16}$O($p,pd$)$^{14}$N$^*$ reaction at $101.3$~MeV is more sensitive to the $pn$ pair amplitude $\hat{\varphi}_{pn}$ than naively expected in the PWIA limit. 
Quantitative extraction of the collectivity through a comparison with experimental data, however, requires a more accurate description of the ($p,pd$) process as mentioned in Sec.~\ref{sec1}. 

\begin{figure}[htbp]
\begin{center}
\includegraphics[width=0.45\textwidth]{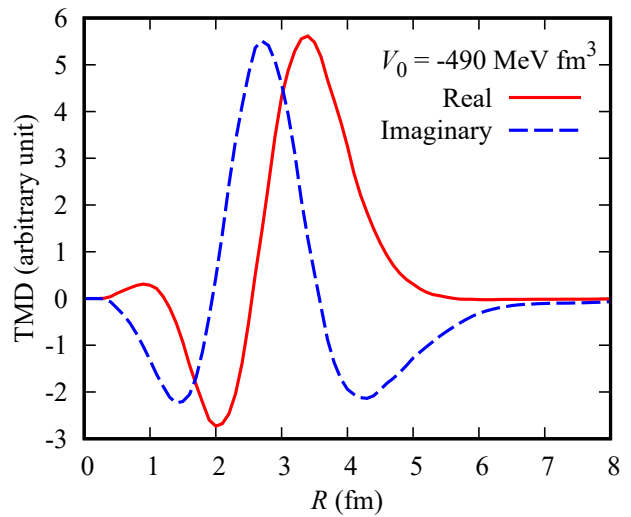}
\caption{Transition matrix density (TMD) corresponding the TDX with $V_0 = -490$~MeV~fm$^3$ at $T^\textrm{L}_1 = 52$~MeV. 
The solid (dashed) line denotes the real (imaginary) part of the TMD.\label{fig6}}
\end{center}
\end{figure}

Figure~\ref{fig6} shows the transition matrix density (TMD) $\delta (R)$, which was originally introduced as a weighting function for the mean density of the ($p,2p$) reaction in Ref.~\cite{KHata97}. 
The definition of the TMD is given by 
\begin{align}
\delta (R) = T^* I (R), 
\label{eq21}
\end{align}
where $I (R)$ is the complex radial amplitude of $T$ of Eq.~(\ref{eq1}), i.e., 
\begin{align}
T = \int^\infty_0 dR~ I (R). 
\label{eq22}
\end{align}
The solid lines denotes the real part of the TMD, which can be interpreted as a radial distribution of the TDX as discussed in Refs.~\cite{KHata97,TNoro99,TWaka17}. 
To make this interpretation plausible, however, the real part of the TMD should not have a large negative value. 
Another condition is that the imaginary part of the TMD is nearly equal to 0 for all $R$. 
As one sees from Fig.~\ref{fig6}, neither of the two conditions is satisfied well. 
This indicates that the interference between amplitudes at different $R$ is strong. 
Furthermore, the TMD is finite even at small $R$, which means the nuclear absorption is not enough to mask the interior region in the evaluation of the transition matrix. 
These features are completely different from for ($p,p\alpha$) reactions discussed in Refs.~\cite{MLyu18,KazuYoshi18}. 
In other words, the distortion effect in the ($p,pd$) reaction investigated in this study is found to be rather complicated and the mechanism for the increase in the relative TDX height due to the distortion is still unclear.

\section{SUMMARY AND PERSPECTIVE}
\label{sec4}
We have investigated the $^{16}$O($p,pd$)$^{14}$N$^*$ reaction at $101.3$~MeV to the $1^+_2$ state of $^{14}$N with the DWIA framework combined with a bound state wave function by EDF. 
As a remarkable feature of the current approach, both the shape and height of the radial wave function of the $pn$ pair in $^{16}$O are evaluated microscopically. 
A clear correspondence between the pairing strength $V_0$ and the TDX was clarified, indicating that the ($p,pd$) reaction is a promising probe for the $T=0$ $pn$ pair in $N \sim Z$ nuclei. 

It is found that the distortion effect enhances the $V_0$ dependence of the TDX. 
Because the selection of the probed region is not clear in the ($p,pd$) process, however, the mechanism of the enhancement is not clear at this stage. 
This is a feature of ($p,pd$) that is quite different from $\alpha$ knockout process, ($p,p\alpha$), in which only the nuclear surface is selectively probed. 

For a quantitative discussion regarding the experimental data, it will be necessary to take into account the deuteron breakup effect in the final channel. 
Another important future work will be the modification of the elementary process of the ($p,pd$) reaction. 
In the current DWIA framework, as in all the preceding DWIA studies, an elastic $p$-$d$ scattering is considered as an elementary process. 
This compels one to assume that a deuteron exists in the target nucleus before the knockout process. 
This may be insufficient to describe the actual ($p,pd$) process, in which a $pn$ pair that is different from deuteron can be knocked out by the incoming proton. 
The pair may form deuteron in the scattering process in the final channel by a coupled-channel effect and then is detected. 
In such a manner, the $p$($pn,d$)$p$ process can be another elementary process for the ($p,pd$) reaction. 
Implementation of both $p$($d,d$)$p$ and $p$($pn,d$)$p$ processes to the coupled-channel DWIA framework will reveal the nature of the $pn$ pair in a nucleus more clearly, and also will be important for applying DWIA to the study of high-momentum $pn$ pair using the backward ($p,pd$) scattering~\cite{Ter18}. 
To achieve this aim, following the ($e,e'd$) analysis \cite{REnt94}, we are constructing a new framework that describes the $p$-$pn$ scattering based on the nucleon degrees of freedom with the nucleon-nucleon effective interaction. 
Studies along these lines are ongoing and will be reported elsewhere. 
For more detailed research, we desire experiments of the ($p,pd$) reaction at higher energy where the DWIA will be able to describe knockout processes with less uncertainty.

\section*{ACKNOWLEDGEMENTS}
This work has been supported in part by Grants-in-Aid of the Japan Society for the Promotion of Science (Grant No. JP16K05352 and JP19K03824). 
The numerical calculations were performed on the CRAY XC40 at YITP, Kyoto University, and on the SX-8 at RCNP, Osaka University. 
The authors acknowledge T.~Uesaka for fruitful discussions.

\appendix
\section*{Appendix: Surface sensitivity of TDX}

\begin{figure}[htbp]
\begin{center}
\includegraphics[width=0.45\textwidth]{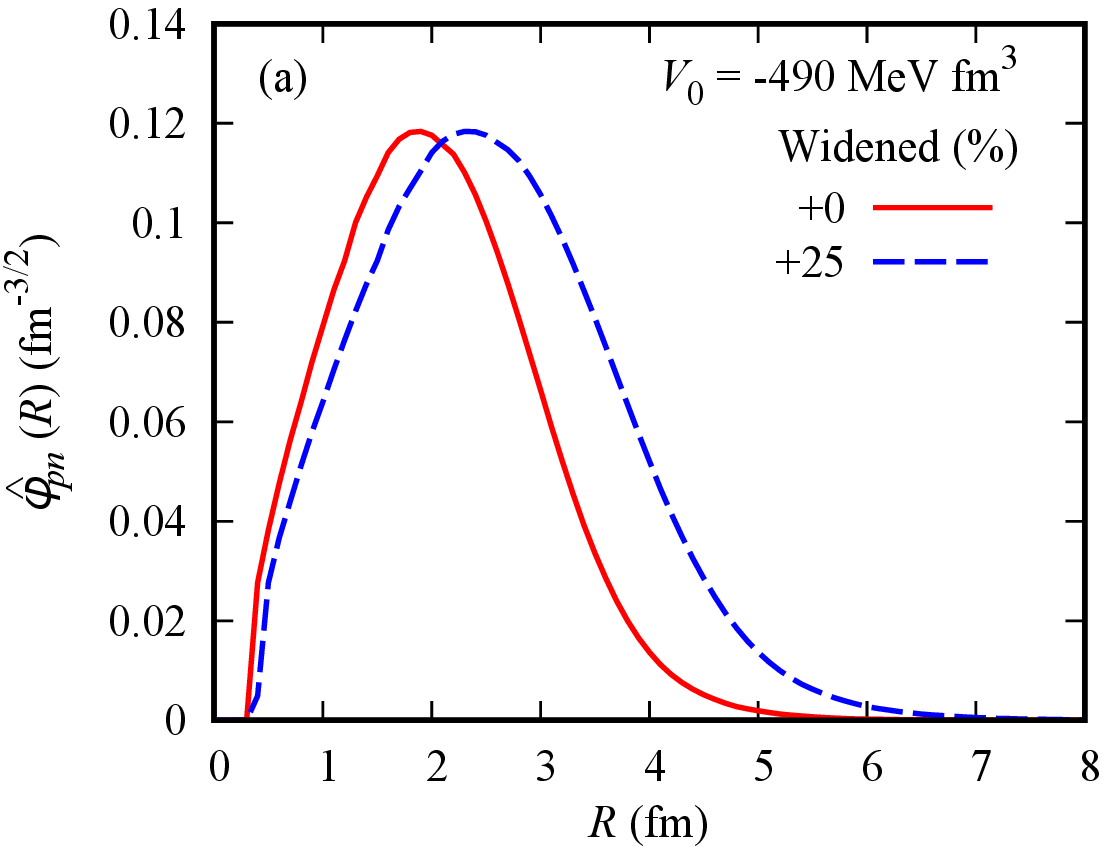}
\includegraphics[width=0.45\textwidth]{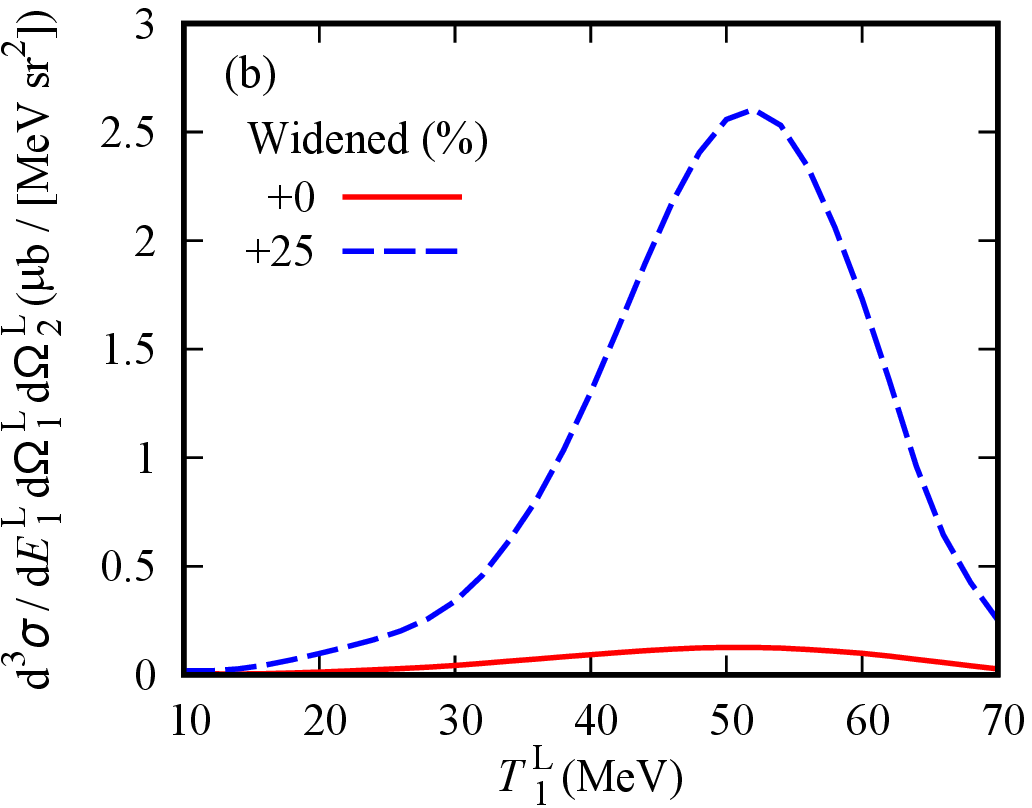}
\caption{(a) The solid line is the same as the dashed line in Fig.~\ref{fig3} divided by $R^2$. 
The dashed line is the wave function obtained by widening the solid line outward by 25{\%} artificially. 
(b) The corresponding TDXs for the $^{16}$O($p,pd$)$^{14}$N$^*$ reaction at $101.3$~MeV.}
\label{fig7}
\end{center}
\end{figure}

Comparing the TDXs shown in Fig.~\ref{fig4} with the experimental data in Ref.~\cite{CSam82}, there is an undershooting by about two orders of magnitude. 
Among the possible sources of this issue, in this Appendix, we show the sensitivity of the TDX to $\varphi_{pn}$ by modifying $\hat{\varphi}_{pn} (R)$ in Eq.~\eqref{eq18}. 

In Fig.~\ref{fig7}(a), the solid line shows the original $\hat{\varphi}_{pn} (R)$ obtained with $V_0 = -490$~MeV~fm$^3$, whereas the dashed line is the result obtained by widening the solid line outward by 25{\%} artificially. 
The TDXs represented by solid and dashed lines in Fig~\ref{fig7}(b) are calculated with the wave functions in the same line type in Fig.~\ref{fig7}(a). 
One can find that the dashed line is about 20 times larger than solid one. 
It means that the 25{\%} extension in the radial distribution of the wave function makes the TDX larger by about a factor of 20 in this case. 

We emphasize that in the actual calculation, the pairing strength $V_0$ is the only variable parameter in the structure model adopted and Fig.~\ref{fig7}(a) does not directly show the uncertainty of the structure model. 
It should be also noted that, as mentioned in Sec.~\ref{sec4}, the improvement in the elementary process of the ($p,pd$) reaction is needed for quantitative comparison with the experimental data. 
It is important, however, to keep in mind how the TDX is sensitive to the radial distribution of the wave function, as demonstrated in this Appendix.


\end{document}